# AOLI: Near-diffraction limited imaging in the visible on large ground-based telescopes.

Craig Mackay*[a], Rafael Rebolo[b,e], David L. King[a], Lucas Labadie[d], Marta Puga Antolin[b], Antonio Pérez Garrido[c], Carlos Colodro-Conde[c], Roberto L. López[b], Balaji Muthusubramanian[d], Alejandro Oscoz[b], J. Rodríguez-Ramos[f], Luis F. Rodríguez-Ramos[b], J. J. Fernandez-Valdivia[f], Sergio Velasco[b]

[a]Institute of Astronomy, University of Cambridge, Madingley Road, Cambridge CB3 0HA, UK
[b]Instituto de Astrofisica de Canarias, C/ Via Lactea s/n, La Laguna, Tenerife E-38205, Spain and Departamento de Astrofísica, Universidad de La Laguna, La Laguna, Spain
[c]Universidad Politecnica de Cartagena, Campus Muralla del Mar, Cartagena, Murcia E-30202, Spain
[d]I. Physikalsiches Institut, Universität zu Köln, Zülpicher Strasse 77, 50937 Köln, Germany
[e]Consejo Superior de Investigaciones Científicas, Spain

## ABSTRACT

The combination of Lucky Imaging with a low order adaptive optics system was demonstrated very successfully on the Palomar 5m telescope nearly 10 years ago. It is still the only system to give such high-resolution images in the visible or near infrared on ground-based telescope of faint astronomical targets. The development of AOLI for deployment initially on the WHT 4.2 m telescope in La Palma, Canary Islands, will be described in this paper. In particular, we will look at the design and status of our low order curvature wavefront sensor which has been somewhat simplified to make it more efficient, ensuring coverage over much of the sky with natural guide stars as reference object. AOLI uses optically butted electron multiplying CCDs to give an imaging array of 2000 x 2000 pixels.

**Keywords:** Lucky Imaging, adaptive optics, Charge coupled devices, EMCCDs, curvature wavefront sensors.

## 1. INTRODUCTION

The angular resolution delivered by ground-based telescopes has changed very little over the last hundred years. Atmospheric turbulence limits resolution to ~1 arcsec on the best ground-based sites. However, the demand from astronomers for better resolution is very strong, and the performance of the Hubble Space Telescope (HST) has had a dramatic effect on the development of many branches of astronomy. Hubble cannot be expected to last forever and increasingly astronomers will look to instrument builders to provide systems that can deliver much sharper images on ground-based telescopes. A great deal of work is currently going into the use of adaptive optic systems to achieve this but most progress has been made in the near infrared on large (4 -10 m class) telescopes. Lucky Imaging is a technique originally suggested by Hufnagel[1] in 1966 and given its name by Fried[2] in 1978. Images are recorded at high frame rates to freeze the motion caused by atmospheric turbulence. The location and quality of each image of each frame allows the best fraction to be selected then shifted and added to give a composite image close to the diffraction limit of the telescope. With a 2.5 m (Hubble sized) telescope then Hubble resolution may be obtained with ~5-30% selection in the visible[3].

Electron multiplying CCDs (EMCCDs) manufactured by E2V Technologies Ltd (Chelmsford, UK) made Lucky Imaging viable as a low light level detection technique. EMCCDs have many of the highly desirable attributes of conventional CCDs. In addition, they have an internal multiplication feature that allows amplification by large factors so that the readout noise which normally limits CCD performance and high frame rates can be made negligible. Even at 30 MHz pixel rate the gain allows individual photons to be detected with good signal-to-noise. This makes these devices particularly well-suited to many high-speed imaging and spectroscopy applications in astronomy and other research areas.

*cdm <at> ast.cam.ac.uk

In principle, telescopes larger than Hubble can also deliver images much sharper than if limited by atmospheric seeing. However, the probability of the Lucky Imaging technique delivering near diffraction limited images becomes vanishingly small for telescopes significantly larger than the HST[2]. This is because the number of turbulent cells across the diameter of the telescope is too large for there to be a significant chance of a relatively flat wavefront (and hence a near diffraction-limited image) across the aperture of the telescope. By increasing the diameter of the telescope we bring in the effects of yet larger scales of atmospheric turbulence. In principle, if we could eliminate the largest turbulent scales where most of the power in the atmospheric turbulence resides[4] then the probability of recording a sharp image will increase. Essentially, eliminating one turbulent scale reduces the phase variance across that scale so that the characteristic cell size, $r_0$ (defined as the scale size over which the variance is ~1 radian$^2$) is increased. Provided enough of the large turbulent scales are removed, the corrected $r_0$ will be large enough so the number of cells across the diameter of the telescope is similar to those typically encountered with an uncorrected 2.4 m aperture.

This method was demonstrated on the Palomar 5-metre telescope behind the PALMAO[5] low order adaptive optics system. Images were obtained with a resolution more than 3 times that of Hubble. In figure 1 we compare our images with those obtained with the Hubble Advanced Camera for Surveys (ACS) with a simple standard lucky selection procedure using the best 10% of the images. Image processing improvements suggested by Garrel[16] et al have been implemented[17] and shown to give greater improvement in the selection percentage that may be used. Further work by Schodel et al[18] has given excellent and more consistent and repeatable results. These methods allow very high selection percentages as shown in that figure. Lucky Imaging has now been used by a number of groups both in Europe and in the US and this has contributed to the scientific development of Lucky Imaging in the visible[6,7,8,9]. For example, FastCam and AstraLux have worked on 2-m and 4-m class telescopes achieving full diffraction-limited resolution in the z'- and I- bands. In the US, projects aiming at developing visible AO systems are now producing very high-resolution images on bright targets[10].

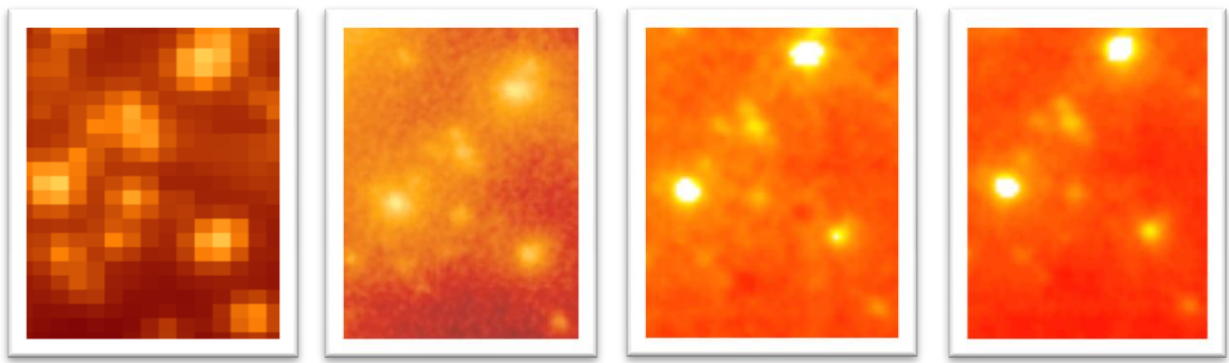

Figure 1: Comparison images of the core of the globular cluster M 13. On the left from the Hubble Advanced Camera for Surveys with ~120 mas resolution followed by an image with the Lucky Camera plus Low-Order AO image with 35 milliarcsecond resolution using the best 10% image selection. Next is the same image but using 20% selection and the Fourier Lucky method[17] followed by the same image but with 50% selection and the Fourier Lucky method. This makes optimum use of images which are degraded in one dimension but have high-resolution information in others[17,18,19].

PALMAO requires a very bright reference star for the Shack-Hartmann wavefront sensor as is normally the case with such wavefront sensors. Generally they need reference stars of I~12-14 magnitude, and these are very scarce[11]. AO systems try to compensate for every wavefront, no matter how complicated so they also have a very small isoplanatic patch in the visible of only a few arcseconds in diameter[12]. As a result, conventional Shack-Hartmann AO based systems can only be used over a tiny fraction of the sky, < 0.1%. Laser guide stars have problems in delivering images with ~0.1 arcseconds accuracy. The consequence of these considerations is that it is very hard to achieve a resolution better than 0.1 arcseconds even in the near infrared on large telescopes although this has been achieved in a limited number of instances, particularly on very bright targets[10].

Another approach to wavefront sensors was needed. A study by Racine[13] showed that curvature sensors actually deployed on telescopes are significantly more sensitive than Shack-Hartmann sensors particularly when used for relatively low order turbulent correction. Olivier Guyon[14] simulated the performance of pupil plane curvature sensors and showed they ought to give a substantial improvement in sensitivity over Shack-Hartmann sensors so allowing allow operation over a much larger fraction of the sky. Using a simplified version of the Guyon proposal to give good low order correction, we have designed a new instrument, AOLI (Adaptive Optics assisted Lucky Imager), to let us carry out science by combining Lucky Imaging and low order AO over much of the sky. AOLI is a collaboration between the Instituto de Astrofisica de Canarias/Universidad de La Laguna (Tenerife, Spain), the Universidad Politecnica de Cartagena (Spain), Universität zu Köln (Germany), the Isaac Newton Group of Telescopes (La Palma, Spain) and the Institute of Astronomy in the University of Cambridge (UK). This paper describes the current status of the instrument, its optical configuration and performance.

## 2. AOLI: GENERAL CONFIGURATION

The AOLI instrument consists of a Tomographic Pupil Image wavefront sensor (TPI-WS) and a low order adaptive optics wavefront corrector using a deformable mirror to feed a Lucky Imaging camera. It is designed specifically for use on the WHT 4.2-m and the GTC 10.4-m telescopes on La Palma but it could be used on almost any large telescope without major modification.

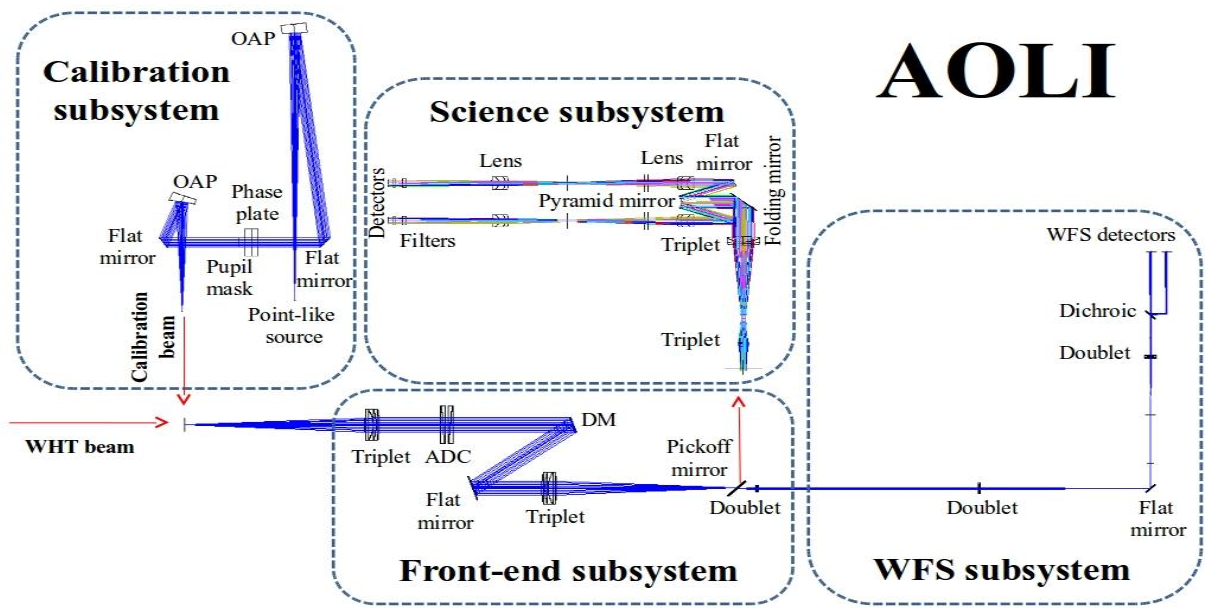

Figure 2: The optical layout of AOLI. Light enters AOLI from the bottom left-hand corner after passing through an image rotator mounted on the Naysmith instrument ring before reaching the telescope focus. The light is collimated and passed through an atmospheric dispersion corrector before it strikes a deformable mirror (DM). We selected the ALPAO-DM241 unit as it has excellent stability and provides an unusually long stroke. The light is then reflected and, after a fold mirror, is reimaged on to a pickoff mirror. The pickoff mirror deflects light towards the science camera (Figure 3). The light from the reference star goes directly on to the wavefront sensor (WFS). The telescope pupil is reimaged down to approximately 2 mm diameter from its original 4.2 m. The wavefront sensor uses a pair of near-pupil images detected on separate EMCCD cameras. In addition, we have a calibration system which is designed to provide calibration sources that are of the same diameter in the telescope focal plane as the diffraction limit of the principal telescope. This is needed particularly to feed the wavefront sensor for setup and alignment. The wavefront sensor uses broadband light between 500 and 950 nm and therefore the calibration system has to be achromatic. Rather than using any refractive optics it uses off-axis paraboloid mirrors.

After the telescope focus we position an atmospheric dispersion corrector, essential for work away from the zenith, after the collimating lens and before the deformable mirror. In Figure 2 the science beam goes from the telescope to a pickoff mirror which deflects light from the science beam towards the science camera shown in figure 3. The reference star is located on the optical axis of the telescope and passes through the pickoff mirror to the wavefront sensor. The light from the telescope is reflected via a deformable mirror set in a pupil plane of the telescope which allows the curvature errors determined by the curvature sensor sub-system to be corrected directly. The deformable mirror is manufactured by ALPAO (France) with 241 elements over the pupil. It will allow correction of wavefront errors on scales of > ~0.5m on the 4.2 m diameter WHT telescope. Our simulations[15] suggest that this will then give us a Lucky Imaging selection percentage under typical/good conditions of about 25-30% in I band. In addition, our simulations suggest that this deformable mirror will have a high enough resolution to achieve satisfactory correction on the GTC 10.4 m telescope. Our approach is to develop a system optimised for the WHT that may be modified and re-deployed quickly in order to demonstrate the technologies as convincingly as possible. Its subsequent deployment to the GTC will then follow.

The science camera (figure 3) is a simple magnifier using custom optics to give diffraction limited performance. The camera is optimised for the 500nm to 1 micron wavelength range. The diffraction limit of the WHT (GTC) 4.2m (10.4 m) telescopes at 0.8μm (I band) is about 40 (15) mas and the camera offers a range of pixel scales of between 12 and 60 mas on the WHT using an interchangeable magnifying lens. The camera uses an array of 4 photon counting, electron multiplying, back illuminated CCD201s manufactured by E2V Technologies Ltd, each 1024 x 1024 pixels. As the CCDs are non-buttable we use an arrangement similar to that of the original HST WF/PC (see Figure 3). Four small contiguous mirrors in the focal plane are slightly tilted and then individually reimaged on to a separate CCD. In order to permit the use of a narrowband filter, for example, for the science object with a broad filter for the reference star, each CCD has its own filter wheel. The configuration allows a contiguous region of 2000 x 2000 pixels giving a field of view of from 120 x 120 arcsec down to 24 x 24 arcseconds on the WHT, depending on the magnification selected.

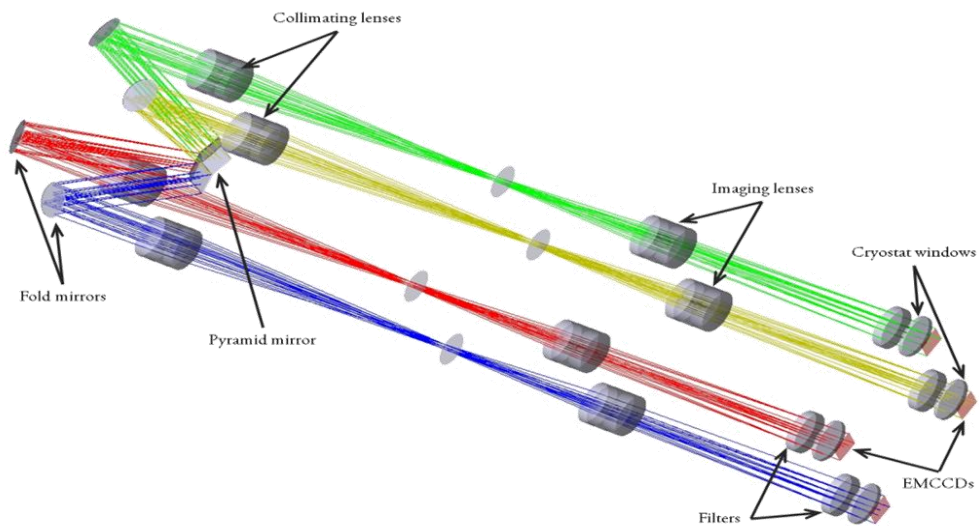

Figure 3: The science camera optical arrangement whereby the light from a single area of sky is split on to four separated and non-buttable CCDs. The magnified image (optics not shown) of the sky is projected onto a pyramid of four mirrors that reflects the light on to relay mirrors and via reimaging optics onto four electron multiplying detectors. This structure was suggested by the design of the original widefield/planetary camera installed on the HST.

The CCDs are back illuminated (thinned) with very high quantum efficiency (peak >95%) from E2V Technologies. Custom electronics developed in Cambridge give up to 30 MHz pixel rate, and 25 frames per sec. With restricted frame format, higher frame rates are possible. For example, a readout format of 2000 x 100 pixels gives ~200 fps, allowing high time resolution astronomy as well with the instrument. The data (~220 MBytes/sec continually) are streamed via the host computer to high-capacity RAID disk drive systems after lossless compression. The host computer performs real-time basic lucky imaging selection, allowing image quality to be assessed while the exposure is progressing. The construction of the unit has been kept relatively low-cost. The instrument will be mounted at Nasmyth focus on an optical bench behind the WHT image rotator. On the GTC it is probable that the instrument would be mounted on one of the folded Cassegrain ports. The layout of AOLI has been modularised because it needs to be transported between the laboratory in La Laguna, Tenerife and the 4.2 m WHT and the 10.4 m GTC telescopes on La Palma[19].

## 3. PHOTON COUNTING CURVATURE WAVEFRONT SENSORS

For the science we wish to carry out it is essential that we are able to use much fainter reference stars that are normally possible with Shack-Hartmann sensors. We only require low order correction and that is something that may be achieved in principle with much fainter reference stars. One form of the curvature wavefront sensor works by taking images on either side of a conjugate pupil plane and by measuring changes in the intensity of illumination as the wavefront passes through the pupil. A part of the wavefront that becomes fainter as it goes through the pupil must correspond to a part of the wavefront that is diverging while if it becomes brighter it is converging. Racine[13] has shown that curvature sensors actually deployed on telescopes are typically 10 times (2.5 magnitudes) more sensitive than Shack-Hartmann sensors for the same degree of correction. In addition they are very much more sensitive again when used for low order correction as the system cell size is dynamically increased[8] and the wavefront sensor readout rate/integration time may be significantly reduced. Correction with coarse cell sizes allows averaging sensor signals over significant areas of the curvature sensor.

One of the key issues when using curvature wavefront sensors is the technique used to achieve an effective fit to the measured wavefronts quickly and reliably. A great deal of work has been put into designing this for Shack-Hartmann sensors where the whole business is very complicated. Essentially it needs high-speed precision matrix inversion to be carried out with negligible latency. With the curvature wavefront sensor things are much more straightforward because the errors produced are simply scaled to match the deformable mirror. The methods that we use are essentially those described by van Dam and Lane[20]. The individual wavefronts processed and analysed using Radon transforms, and the fitting process proceeds very rapidly. The reconstruction methods were developed by a team at the University of La Laguna who had been involved for some time in developing plenoptic[21] wavefront sensors. Details of this may be found in the thesis (in Spanish) by J. J. Fernandez-Valdivia[22]. These techniques, which are used by AOLI, have a number of advantages. It is possible to use a broadband light source and indeed the reference object does not need to be unresolved only that it should be properly representative of the wavefront errors to correct.

AOLI uses two near-pupil planes to provide a rapid convergence in achieving a wavefront fit. The angular resolution we should achieve on the WHT/GTC will be typically 40/15 milliarcseconds in the visible to I-band range. Our simulations suggest that we should be able to operate at the faintest level of the reference star of about 17.0-18.0 on the WHT and about 18.0-19.0 mag on the GTC. This will allow us to find reference stars over nearly all the sky even at high galactic latitudes (>85%)[11]. At the fainter end of these ranges it is probable that we will only be able to achieve partial correction but enough to improve the variance of the wavefront entering the science camera.

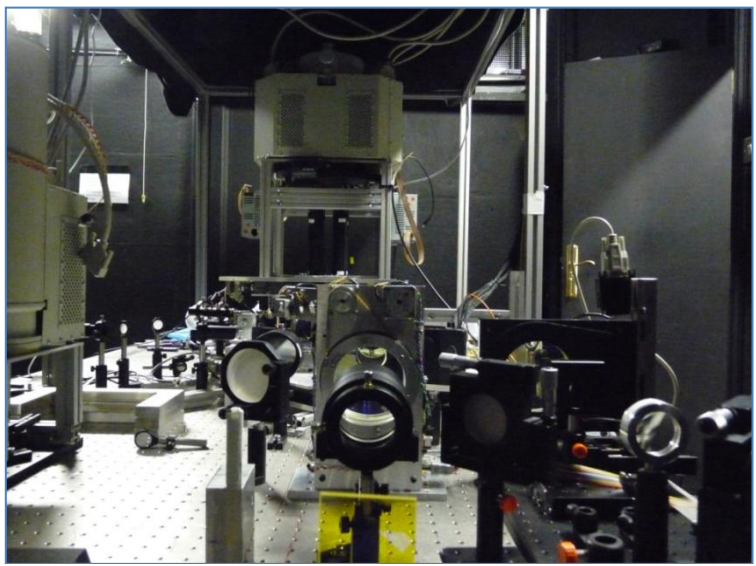

Figure 4: AOLI mounted at the Naysmith focus of the William Herschel 4.2 m Telescope. Most of the light baffles used in the instrument were removed before this photograph was taken.

## 4. FIRST LIGHT EXPERIENCES WITH AOLI

The first observing run with AOLI was on the 4.2 m William Herschel telescope (WHT). In most respects the instrument worked well and figure 4 shows it mounted at the Naysmith focus of the WHT. Unfortunately the weather conditions while at the telescope were remarkably poor with very poor seeing, high humidity and frequent dome closures. As a consequence very little in the way of scientific results were actually obtained (Velasco et al.[23]). A second run in May 2016 also suffered from very poor weather conditions and useful results were obtained during the 2 hours in which we had seeing slightly below 2 arcsec. Most importantly, we were able to demonstrate that the curvature wavefront sensor did indeed work and that the size of the images recorded were dramatically improved by the combination of Lucky Imaging and low order adaptive optics driven by the curvature wavefront sensor. Images of this are shown in figure 5.

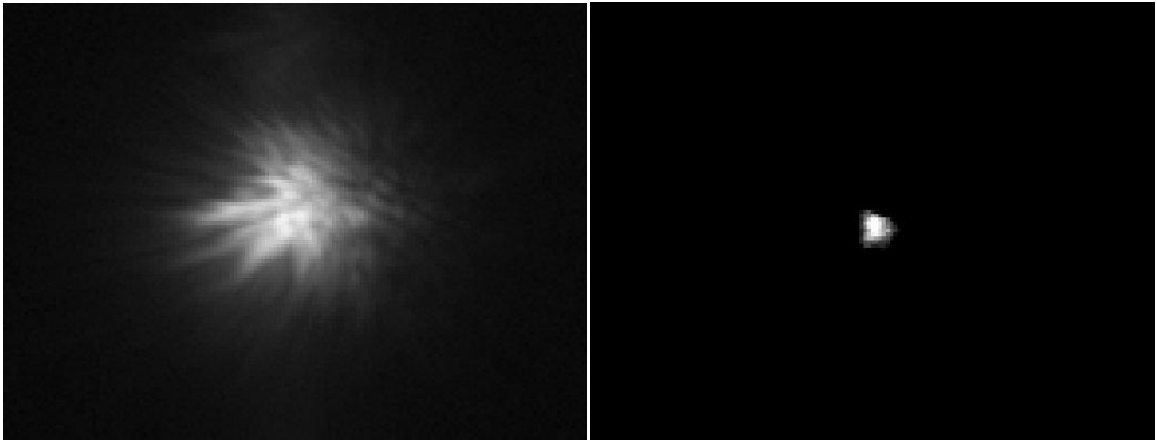

Figure 5: An example of the image delivered by AOLI firstly (left frame) without any attempt at wavefront correction and then (right frame) with the AO-loop closed. Substantial improvement in image resolution is visible clearly

## 5. CONCLUSIONS

AOLI is now close to delivering images at the diffraction limit of a 4 m class telescope in the visible from the ground. This has been the goal for many years of instrument developers, trying to produce a near-diffraction limited instrument capable of observing over the full sky with good observing efficiency in the visible part of the spectrum. The development of EMCCDs with their photon counting availability has transformed our capacity to build instruments capable of working with the rapidly changing atmosphere and using very faint reference stars to let us optimise our recording of light from the sky. The combination of these devices with low order curvature wavefront sensors (also using EMCCDs) will allow a new generation of astronomers to explore the Universe with as big a step change in resolution as Hubble provided over 20 years ago. Hubble provided an eight-fold improvement over the typical ground-based image resolution of ~1 arcsec to give images of ~0.12 arcsec resolution. We have already demonstrated a further improvement over Hubble with the Palomar 5 m telescope by imaging with ~0.035 arcsec resolution. AOLI in the visible on the GTC 10.4m telescope has the diffraction limit of eight times better than HST of ~0.015 arcsec resolution, roughly 60 times the resolution when limited by atmospheric turbulence. There is every expectation that by making such high resolution images and spectra available more routinely many fields of astronomy will be revolutionised yet again. It is important to realise that AOLI has the potential to feed not only an imaging camera but also an integral field spectrograph of other instruments.

# REFERENCES

[1] Hufnagel, R.E., in "Restoration of Atmospheric and Degraded Images" (National Academy Of Sciences, Washington, DC, 1966), vol 3, Appendix 2, p11, (1966).
[2] Fried, D L, "Probability of Getting a Lucky Short-Exposure Image through Turbulence", JOSA, vol 68, p1651-1658,(1978).
[3] C. D. Mackay, T. D. Staley, D. King, F. Suess and K. Weller, "High-speed photon-counting CCD cameras for astronomy", Proc SPIE 7742, p. 1-11,(2010).
[4] Noll, R. J., "Zernike Polynomials and Atmospheric Turbulence", JOSA, vol 66, p207-211, (1976).
[5] N.M. Law, C.D. Mackay, R.G. Dekany, M. Ireland, J. P. Lloyd, A. M. Moore, J.G. Robertson, P. Tuthill, H. Woodruff, "Getting Lucky with Adaptive Optics", ApJ ,vol. 692 p.924-930, (2009).
[6] C. Bergfors, W. Brandner, M. Janson et al. "Lucky Imaging Survey for southern M dwarf binaries", Astron & Astrophys, vol. 520, A54 (2010).
[7] D. Peter, M. Feldt, Th. Henning, and F. Hormuth, "Massive binaries in the Cepheus OB2/3 region – Constraining the formation mechanism of massive stars", Astron. & Astrophys., vol. 538, A74 (2012).
[8] L. Labadie, R. Rebolo, I. Villo et al., "High-contrast optical imaging of companions: the case of the brown dwarf binary HD 130948 BC", Astron. & Astrophys., vol. 526, A144 (2011).
[9] B. Femenia, R. Rebolo, J. A. Perez-Prieto et al., "Lucky Imaging Adaptive Optics of the brown dwarf binary GJ569Bab", MNRAS, vol. 413, p.1524-1536 (2011).
[10] L.M. Close, et al., "Diffraction-Limited Visible Light Images of Orion Trapezium Cluster with the Magellan Adaptive Secondary Adaptive Optics System (MagAO)",ApJ, 774, 94 (2013).
[11] Simons, D., " Longitudinally Averaged R-Band Field Star Counts across the Entire Sky ", Gemini Observatory technical note, (1995).
[12] M. Sarazin and A. Tokovinin, "The Statistics of Isoplanatic Angle and Adaptive Optics Time Constant Derived from DIMM Data", conference "Beyond Conventional Adaptive Optics", Venice (2001).
[13] R. Racine,"The Strehl Efficiency of Adaptive Optics Systems", Pub. Astr. Soc Pacific, vol 118, p1066-1075, (2006).
[14] O. Guyon "Limits of Adaptive Optics for High-Contrast Imaging", Ap J, vol 629,p 592-614,(2005)
[15] J.Crass, C.D. Mackay, et al,"The AOLI low-order non-linear curvature wavefront sensor: a method for high sensitivity wavefront reconstruction", Proc SPIE 8447, (2012).
[16] V. Garrel et al "A Highly Efficient Lucky Imaging Algorithm: Image Synthesis Based on Fourier Amplitude Selection", PASP, 124,861 (2012)
[17] Mackay, C.D., "High-Efficiency Lucky Imaging", MNRAS, 432,702 (2013)
[18] Schodel, R. et al, "Holographic imaging of crowded fields: high angular resolution imaging with excellent quality at very low cost", MNRAS, 429,1367 (2013)
[19] Oscoz, A. et al, "An instrumental puzzle: the modular integration of AOLI", SPIE, 9908-113, June 2016.
[20] van Dam, M & Lane, R.," Extended Analysis of Curvature Sensor", JOSA, 19,1390 (2002)
[21] Rodríguez-Ramos, L. et al., "Concepts, laboratory and telescope test results of the plenoptic camera as a wavefront sensor", SPIE, 8447, (2012).
[22] Valdivia,JJ, "Cofaseado de segmentos y óptica adaptativa con sensor geométrico", PhD thesis, University of La Laguna, December 2015.
[23] Velasco, S., et al. "High spatial resolution optical imaging of the multiple T Tauri system LkHα 262/LkHα 263", MNRAS, 460,3519 (2016)